\documentclass{aimc2026}

\usepackage[utf8]{inputenc} % allow utf-8 input
\usepackage[T1]{fontenc}    % use 8-bit T1 fonts
\usepackage{hyperref}       % hyperlinks
\hypersetup{
  colorlinks = true,  % Colours links instead of ugly boxes
  urlcolor   = blue,  % Colour for external hyperlinks
  linkcolor  = black, % Colour of internal links
  citecolor  = black  % Colour of citations
}
\usepackage{url}            % simple URL typesetting
\usepackage{booktabs}       % professional-quality tables
\usepackage{amsfonts}       % blackboard math symbols
\usepackage{nicefrac}       % compact symbols for 1/2, etc.
\usepackage{microtype}      % microtypography
\usepackage{graphicx}       % figures
\usepackage{cleveref}       % automatic figure references

\usepackage{enumitem}           % Customizable lists (used in itemize settings)

\usepackage{pifont}
\usepackage{xspace}
\usepackage{multirow}
\usepackage{array}
\newcolumntype{L}[1]{>{\raggedright\arraybackslash}p{#1}}
\newcommand{\fully}{\ding{51}\xspace} % ✓
\newcommand{\notsupported}{\ding{55}\xspace} % ✗
\newcommand{\partially}{$\sim$\xspace}

\usepackage{xurl}
\title{MusGU+: Toward a Musician-Centered Evaluation Framework and Discovery Tool for Generative Music AI}

\author{%
  Laura Ibáñez-Martínez\\
  Music Technology Group\\
  Universitat Pompeu Fabra\\
  Barcelona, Spain \\
  \texttt{laura.ibanez@upf.edu}\\
  \And
  Roser Batlle-Roca\\
  Music Technology Group\\
  Universitat Pompeu Fabra\\
  Barcelona, Spain\\
  \texttt{roser.batlle@upf.edu}\\
  \AND
  Xavier Serra\\
  Music Technology Group\\
  Universitat Pompeu Fabra\\
  Barcelona, Spain\\
  \texttt{xavier.serra@upf.edu}\\
  \And
  Martín Rocamora\\
  Music Technology Group\\
  Universitat Pompeu Fabra\\
  Barcelona, Spain\\
  \texttt{martin.rocamora@upf.edu}
}

\begin{document}

%TC:ignore
\twocolumn[%
  \maketitle
\begin{abstract}
  Generative music systems are increasingly presented as tools that democratize music creation, yet their practical suitability for musicians remains underexplored. Prior work includes openness-focused evaluation frameworks, such as MusGO (Music-Generative Open AI), as well as qualitative studies of musicians' experiences with generative systems. However, these approaches do not support systematic comparison or early-stage discovery of models for creative use. Motivated by such limitations, we introduce MusGU+, a musician-centered evaluation framework organized around three dimensions: Adaptability, Usability, and Controllability. Together, these capture whether a model can be feasibly trained or fine-tuned on personal data, integrated into real-world music workflows, and controlled in musically meaningful ways. We evaluate 10 representative generative music systems and present an interactive discovery tool that enables musicians to explore and filter models according to these criteria. While MusGO remains valuable for promoting responsible research practices, MusGU+ supports informed selection and practical adoption of generative systems by musicians.
\end{abstract}%
]
%TC:endignore

\aimcnotice % name of proceedings title in the first column bottom

\section{Introduction}

Generative music systems have rapidly gained popularity in recent years, particularly through consumer-facing platforms such as Suno \citep{suno2026} and Udio \citep{udio2026}.
At the same time, these systems have attracted criticism for their limited transparency regarding training data \citep{reuters2024sunolawsuit}, as well as broader ethical concerns related to authorship, creative agency, and environmental impact \citep{barnett2023ethical}, contributing to growing skepticism toward generative AI.
Despite these concerns, such platforms are frequently presented as democratizing music creation \citep{pram2025opening}, which raises a fundamental question: \textit{for whom are these systems actually intended?}
While platforms like Suno and Udio lower technical barriers and offer more user-friendly interfaces than earlier generative systems, their underlying assumptions about music-making remain at odds with how many musicians understand and value their creative practice.
For instance, the CEO of Suno has stated: “the majority of people don’t enjoy the majority of the time they spend making music” \citep{djmag2025suno}, framing generative systems as technologies for optimizing music creation rather than fostering creativity and artistic exploration.

Existing evaluation approaches for generative music systems have mostly relied on output-level evaluation \citep{agostinelli2023musiclm, copet2024simple, evans2025stableaudio} and, more recently, representation-level analysis \citep{wei2024syntheory, ibanez2026disentanglement}, offering limited insight into how systems can be adapted, controlled, or integrated into real-world musical contexts \citep{lerch2025survey}.
Complementary approaches have begun to address these limitations from different perspectives.
In particular, the Music-Generative Open AI (MusGO) framework \citep{batlleroca2025musgo} assesses systems in terms of openness and reproducibility.
Other work has examined musician-relevant considerations such as data provenance, licensing, creative agency, and user control \citep{wilson2025responsibleai}, while practice-oriented studies have further explored how such systems can be incorporated into creative workflows \citep{dadman2025workflow, ronchini2025aiassisted}.
However, these approaches do not provide a systematic comparison of the practical characteristics that shape the suitability of generative systems for musicians, particularly in terms of adaptation, control, and workflow integration.

In this work, we introduce \textbf{Music-Generative Usable+ AI (MusGU+)}, a musician-centered evaluation framework designed to address these limitations.
MusGU+ adopts the composite and graded evaluation approach followed in MusGO, while shifting the emphasis toward practical suitability for musicians.
In particular, it focuses on whether models can be adapted to personal data, integrated into real-world music workflows, and controlled in musically meaningful ways.
The framework structures evaluation around three dimensions: \textbf{Adaptability}, \textbf{Usability}, and \textbf{Controllability}, capturing factors such as training pathways, interface availability, workflow integration, and control mechanisms.
We apply MusGU+ to 10 generative music systems and present the results through an interactive discovery tool\footnote{\url{https://lauraibnz.github.io/MusGU-plus/}} that enables musicians to explore, compare, and filter models according to the framework’s evaluation criteria\footnote{\url{https://lauraibnz.github.io/MusGU-plus/framework.html}}, supporting more informed selection and practical adoption in creative contexts.
As with MusGO, the MusGU+ repository\footnote{\url{https://github.com/lauraibnz/MusGU-plus}} is intended as a living resource, encouraging community engagement through ongoing updates, model additions, and discussion.
More broadly, we hope our work encourages further consideration of musician-centered aspects in the development of new generative music systems.

\section{Background and Related Work}

\subsection{Evaluating Generative Music AI}

Evaluation of generative music AI has predominantly relied on output-level metrics, particularly for model development and benchmarking.
Representative text-to-music systems such as MusicLM \citep{agostinelli2023musiclm}, MusicGen \citep{copet2024simple}, and Stable Audio Open \citep{evans2025stableaudio} typically assess generation quality using objective metrics such as Fréchet Audio Distance (FAD) \citep{kilgour2019fad} and Kullback-Leibler (KL) divergence \citep{kreuk2022audiogen}, alongside CLAP-based scores \citep{wu2023clap} for text-prompt adherence.
These are often complemented by perceptual listening tests, where participants assess generated samples in terms of realism, coherence, stylistic consistency, and overall preference.

Beyond output-focused evaluation, recent work has begun to analyze the internal representations learned by generative music systems.
Such analyses aim to uncover what musical attributes are encoded within these representations, and how this knowledge might support interpretability and controllability.
Examples include studies that probe learned representations through music-theoretic lenses \citep{wei2024syntheory} and approaches that pursue disentanglement to enable controllable music generation \citep{ibanez2026disentanglement}.
More recent work has also applied mechanistic interpretability methods to identify and manipulate interpretable musical concepts within generative models \citep{singh2026discovering}.
These approaches provide insight into model behavior that is not directly observable from outputs alone.

While output- and representation-level evaluations offer important perspectives on quality and model behavior, they remain limited in accounting for how systems are used and experienced in practice.
As noted by \citet{lerch2025survey}, strong performance on objective metrics or favorable perceptual scores does not necessarily correspond to systems that are usable or controllable in real musical contexts.

\subsection{Openness-Focused Evaluation}

Among evaluation approaches that consider generative music systems as a whole, a growing focus has been placed on their openness, including source code availability, training data disclosure, licensing, and documentation.
The MusGO framework \citep{batlleroca2025musgo} is a community-driven initiative that assesses generative systems across these dimensions, with the aim of promoting transparency and reproducibility.
It also addresses cases of “open-washing”, in which models are presented as open despite releasing only a limited subset of components \citep{liesenfeld2024}.
MusGO provides a comparative overview through an openness leaderboard, supporting ongoing updates and community contributions.

While MusGO acknowledges factors that are relevant to musical practice, such as computational requirements and user-oriented applications, these are not the primary focus of the evaluation.
Consequently, models with substantially different levels of practical accessibility or usability may appear similar from an openness perspective.
This risks obscuring meaningful differences in the conditions under which systems can be adopted and integrated into real-world musical workflows.

A related perspective grounded in responsible AI principles comes from \citet{wilson2025responsibleai}, who survey contemporary generative music systems across several musician-relevant dimensions, including input type, real-time generation, licensing, and training or fine-tuning options.
While this work raises critical ethical and practical questions, it is primarily a descriptive analysis of the current landscape rather than a structured evaluation.
Its focus on musician-relevant concerns highlights considerations that are not systematically reflected in openness-focused evaluations.

\subsection{Toward Musician-Centered Perspectives}

In parallel, a growing body of literature has examined generative music systems as tools within creative practice, rather than as abstract generators.
Several studies have focused on text-to-music models, embedding them in music production settings to study how they support ideation and practical use.
\citet{ronchini2025aiassisted} report that while text-to-music systems enable rapid exploration, musicians often struggle to translate musical intent into effective control through textual prompts, particularly with respect to temporal structure, arrangement, and expressive detail.
Related work \citep{choi2025promptbased} similarly highlights mismatches between linguistic descriptions and the time-varying, multidimensional nature of musical control.

Further insight comes from workflow-oriented evaluations of generative systems in sound design and music production.
\citet{dadman2025workflow} examine a range of models through hands-on experimentation, identifying challenges related to steerability, latency, configuration complexity, and integration into iterative creative processes.
Studies with professional sound designers likewise emphasize the importance of fine-grained and predictable control, noting that opaque or coarse control mechanisms constrain expressive precision and sustained creative use \citep{kamath2024sounddesigner}.
Beyond interaction and control, \citet{bryankinns2025smalldatasets} examine the role of data scale and system design, highlighting limitations of large-scale training paradigms for non-mainstream musical practices and the potential of small-data approaches to support creative agency.

Other studies adopt a more observational perspective, examining how musicians already incorporate generative AI into their workflows.
\citet{pons2025musicai} document a diverse landscape of AI-assisted music-making across applications such as music production, sound design, and live performance.
Their analysis shows that musicians often employ generative models as components within broader creative pipelines, and that working with such systems can be technically demanding, often requiring technical expertise or community support.
Thus, despite their growing availability, integration of generative systems into creative workflows remains shaped by both interaction and technical constraints, while existing work offers limited support for systematically comparing models according to musicians' practical needs, particularly in early stages of adoption.

\section{MusGU+: A Musician-Centered Evaluation Framework}

\subsection{Framework Overview and Design Principles}

We introduce the \textbf{Music-Generative Usable+ AI (MusGU+)} framework, which adopts a composite and graded evaluation approach inspired by MusGO \citep{batlleroca2025musgo} and prior openness-focused literature \citep{liesenfeld2024}.
Rather than treating evaluation criteria as binary properties, this approach characterizes systems along multiple dimensions, each assessed through clearly defined graded criteria.
While MusGO focuses on openness and responsible research practices, MusGU+ shifts toward practical suitability for creative use.
Like MusGO, MusGU+ is conceived as a community-oriented living framework, intended to support further refinement, the evaluation of new models, and continued discussion around musician-centered concerns.

The design of MusGU+ draws on prior literature and the authors’ experience as researchers in music technology, with a particular focus on generative music systems, including work on controllability and ethical aspects of AI. It also builds on their practice as musicians across different musical contexts, including instrumental performance, music production, and involvement in electronic music scenes.

The framework was iteratively developed through internal discussion and cross-review, and further refined through informal consultation with music producers and sound designers, with the aim of ensuring that its concepts and terminology are understandable and meaningful for practitioners.
It was subsequently discussed with participants in the 5th Generative Music AI Workshop\footnote{\url{https://www.upf.edu/web/mtg/generative-music-ai-workshop}}, held at the Music Technology Group (MTG), providing an additional opportunity for feedback.
In this sense, MusGU+ retains a systematic evaluation structure while foregrounding characteristics relevant to musicians and their creative practice.

\subsection{Defining Musician-Centered Axes}

One motivation for articulating musician-centered evaluation axes emerged from limitations identified in openness-focused criteria, which do not explicitly account for certain practical aspects of model use.
For example, RAVE \citep{caillon2021rave} supports training on small, user-curated datasets, facilitating adaptation to musicians’ own material.
While such characteristics are not directly reflected in openness-focused criteria, they have been central to RAVE's use in practice.
This aligns with prior work highlighting limitations of large-scale training paradigms and the potential of small-data approaches to support creative agency \citep{bryankinns2025smalldatasets}.
Together, these considerations motivate evaluating the feasibility and available options for adapting models to musicians' own data.

Musician-centered evaluation also requires considering the conditions under which generative systems can be accessed and integrated into musical workflows.
RAVE, for instance, supports real-time interaction through environments such as Max/MSP and plugin-based systems \citep{wilson2025responsibleai, dadman2025workflow}.
More broadly, prior work highlights technical and workflow constraints that can shape the practical use of generative systems \citep{dadman2025workflow, pons2025musicai}.
This suggests that practical accessibility depends not only on system availability, but also on how readily systems can be run, interacted with, and integrated into creative workflows.

A further motivation concerns controllability, particularly the gap between available control mechanisms and musically meaningful guidance.
Prompt-based systems offer limited control over temporal structure and expressive detail \citep{ronchini2025aiassisted, choi2025promptbased}.
By contrast, recent approaches such as AFTER \citep{demerle2024combining}, along with earlier work like DDSP \citep{engel2020ddsp}, provide more explicit control over musical attributes, improving interpretability and controllability \citep{ibanez2026disentanglement}.
In addition, RAVE \citep{caillon2021rave} and AFTER \citep{demerle2024combining} expose internal representations through interactive interfaces, supporting exploratory interaction.
These examples highlight the relevance of control mechanisms and interaction possibilities that may support different forms of creative use.

Building on these insights, we organize MusGU+ around three musician-centered evaluation axes:

\vspace{-0.5em}
\begin{itemize}[label={}, leftmargin=0.5em]
    \item \textbf{Adaptability.} Assesses how feasible it is for musicians to adapt a model to their own data, including practical constraints and supported training or fine-tuning pathways.
    \item \textbf{Usability.} Examines how easily models can be run, interacted with, and integrated into creative workflows, including access constraints and available support channels.
    \item \textbf{Controllability.} Evaluates the input options and control mechanisms provided by the model, including the diversity, structure, and independence of its control pathways.
\end{itemize}

\subsection{Evaluation Criteria and Structure}

The three musician-centered axes are operationalized through 15 evaluation criteria: Adaptability (5), Usability (6), and Controllability (4).
A detailed description of the evaluation criteria is provided in Appendix~\ref{app:criteria}.

For Adaptability, these criteria include \textit{Hardware Requirements}, \textit{Dataset Size}, \textit{Adaptation Pathways}, \textit{Technical Barriers}, and \textit{Model Redistribution}.
Together, these criteria reflect whether a system can be adapted by musicians using realistic computational resources and personal datasets, whether suitable training or fine-tuning pathways are provided, and whether adapted models can be reused or shared beyond the original system.

Usability is evaluated through the criteria \textit{Interface Availability}, \textit{Access Restrictions}, \textit{Real-Time Capabilities}, \textit{Workflow Integration}, \textit{Output Licensing}, and \textit{Community Support}.
These criteria assess whether a system can be accessed and run reliably, used interactively, and incorporated into music production or sound design workflows.

Controllability, in turn, includes the criteria \textit{Conditioning Inputs}, \textit{Time-Varying Control}, \textit{Feature Disentanglement}, and \textit{Control Parameters}.
These criteria aim to capture the control mechanisms available to musicians and their potential to support intentional musical decisions through precise and predictable guidance of the generation process.

Each criterion is assessed using a three-level graded scale: \textit{fully} (\fully), \textit{partially} (\partially), and \textit{not supported} (\notsupported), according to the degree to which the corresponding capability is satisfied.
In each case, the score levels are accompanied by short descriptive explanations specifying the conditions under which a system would fall into a given category, including representative examples where applicable.

\begin{table*}[t]
\centering
\setlength{\tabcolsep}{4pt}
\renewcommand{\arraystretch}{1.5}
\caption{Overview of evaluated generative music systems and rationale for their inclusion in relation to MusGU+ axes.}
\label{tab:model_selection}
\small
\begin{tabular}{L{3.2cm}L{0.7cm}L{2.1cm}L{2.5cm}L{7.11cm}}
\toprule
\textbf{Model} & \textbf{Year} & \textbf{Architecture} & \textbf{Context} & \textbf{Rationale for Inclusion} \\
\midrule
AFTER & 2024 & Latent diffusion, rectified flow & Academic &
\textbf{Adaptability:} Pretrained audio codec. \newline
\textbf{Usability:} Real-time; Max/MSP, Max4Live. \newline
\textbf{Controllability:} Disentangled, time-varying control. \\
DDSP-VST & 2022 & Differentiable DSP & Industrial Research &
\textbf{Adaptability:} Small datasets; custom model training. \newline
\textbf{Usability:} Real-time; VST/AU. \newline
\textbf{Controllability:} Time-varying, interpretable control. \\
JAM & 2025 & Latent diffusion, rectified flow & Academic &
\textbf{Adaptability:} ``Tiny'' architecture. \newline
\textbf{Controllability:} Word- and phoneme-level timing control. \\
MusicGen & 2023 & Autoregressive transformer & Industrial Research &
\textbf{Adaptability:} Training and fine-tuning pathways. \newline
\textbf{Controllability:} ``Controllable''; time-varying melody conditioning. \\
Neutone Morpho & 2026 & Autoencoder & Proprietary &
\textbf{Adaptability:} Low-barrier custom model training. \newline
\textbf{Usability:} Real-time; VST/AU. \\
RAVE & 2021 & Variational autoencoder & Academic &
\textbf{Adaptability:} Small datasets; custom model training. \newline
\textbf{Usability:} Real-time; VST/AU, Max/MSP. \newline
\textbf{Controllability:} Latent space exploration. \\
Stable Audio Open Small & 2025 & Latent diffusion & Industrial Research &
\textbf{Usability:} ``Fast'' text-to-audio generation. \\
Suno & 2026 & -- & Proprietary &
\textbf{Usability:} Web interface; active Discord community. \\
Udio & 2024 & -- & Proprietary &
\textbf{Usability:} Web interface; active Discord community. \\
YuE & 2025 & Autoregressive transformer & Academic &
\textbf{Controllability:} Section-based lyrics-to-song conditioning. \\
\bottomrule
\end{tabular}
\end{table*}

\section{Model Evaluation with MusGU+}

\subsection{Model Selection}

We apply the MusGU+ framework to 10 generative music systems in the audio domain:
AFTER \citep{demerle2024combining}, DDSP-VST \citep{ddspvst2026}, JAM \citep{liu2025jam}, MusicGen \citep{wu2024}, Neutone Morpho \citep{neutone2026}, RAVE \citep{caillon2021rave}, Stable Audio Open Small \citep{evans2025stableaudio}, Suno \citep{suno2026}, Udio \citep{udio2026}, and YuE \citep{yuan2025yue}.
An overview of these models is provided in Table~\ref{tab:model_selection}.

This selection aims to capture a range of current generative music systems that exhibit characteristics relevant to one or more MusGU+ axes and can plausibly be incorporated into real-world music-making workflows.
Consequently, previously influential models that are no longer actively maintained, such as Jukebox \citep{dhariwal2020jukebox} and Musika \citep{pasini2022musika}, are excluded, despite their significance at the time of release.
This choice reflects MusGU+'s intention to support early-stage discovery and informed adoption, rather than retrospective evaluation.

The selected models also span diverse architectures and development contexts, covering both academic research and industrial or product-oriented environments, enabling comparison across design goals, intended uses, and musical applications.
While not exhaustive, the sample reflects the types of systems MusGU+ is designed to evaluate.

\subsection{Evaluation Process}

We follow an iterative evaluation process inspired by the methodology used in MusGO.
Each model is first evaluated by one of the authors through a review of official resources provided by the developers, including research papers, project websites, documentation, code repositories, and released interfaces.
These materials are used to assign scores for each criterion, accompanied by a short justification for every assessment.
We also consider supplementary resources acknowledged within the project ecosystem, such as community-developed training notebooks, auxiliary interfaces, or third-party tools referenced by the model developers.
Available user interfaces are inspected to assess their functionality and integration pathways, although perceptual or qualitative evaluation of model behavior and generated audio is beyond the scope of this work.

Each evaluation is then independently reviewed by another author.
Discrepancies are discussed and resolved through joint inspection of the evidence, and scores and justifications are refined through one or more iterations until consensus is reached.
This cross-review process aims to improve consistency, reduce individual bias, and ensure that evaluations are grounded in verifiable information.

\subsection{Interactive Display of Results}

\begin{figure*}[ht]
  \centering
  \includegraphics[width=\linewidth]{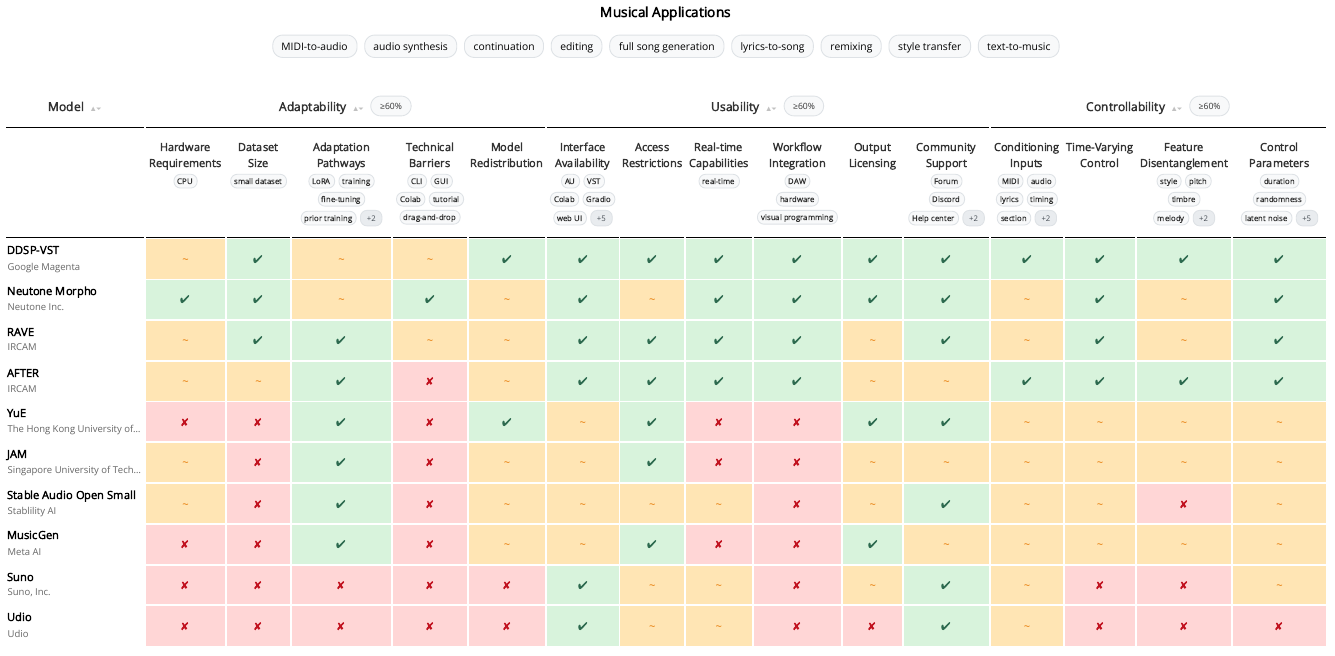}
  \caption{Interactive display of MusGU+ evaluation results for the selected generative music systems across Adaptability, Usability, and Controllability, enabling filtering, reordering, and comparison according to musical needs and priorities.}
  \label{fig:musgu_eval}
\end{figure*}

The results of the MusGU+ evaluation are shown in Figure~\ref{fig:musgu_eval}.
Rather than using a static table or leaderboard, we adopt an interactive display that supports exploration and comparison across musical needs and creative priorities.
This design choice reflects the view that no single ranking can meaningfully capture the suitability of generative systems across diverse musical contexts, and that evaluation outcomes are most informative when they can be inspected, reordered, and filtered according to intended use.

Aggregate scores are computed for each model by assigning values of 0 for \notsupported (\textit{not supported}), 0.5 for \partially (\textit{partially supported}), and 1 for \fully (\textit{fully supported}), and summing across criteria and axes.
While this aggregation provides a default ordering, it is not intended as a definitive ranking.
Instead, the interface allows reordering by individual axes and filtering based on minimum score thresholds (e.g., models with at least 60\% support within a given axis).

While aggregate scores provide a useful overview, they do not capture important qualitative distinctions between systems.
We therefore introduce reusable tags for selected criteria to support more interpretable and task-oriented comparison.
For example, within \textit{Conditioning Inputs}, tags such as \textit{audio}, \textit{MIDI}, and \textit{text prompt} indicate supported input modalities, while within \textit{Workflow Integration}, tags such as \textit{DAW} and \textit{hardware} capture how systems align with established production and performance environments.
For other criteria, tags correspond more directly to score levels.
In the case of \textit{Real-Time Capabilities}, a single \textit{real-time} tag represents systems that achieve \textit{fully supported} (\fully), indicating support for low-latency or interactive use.

The tagging scheme was developed iteratively during the evaluation process, informed by analysis of the selected systems and recurring discussion among the authors.
In some cases, this involved abstracting across heterogeneous terminology.
For instance, within \textit{Control Parameters}, technical controls (e.g., \textit{temperature}) and more abstract concepts from proprietary systems (e.g., \textit{serendipity}) are mapped to a shared \textit{randomness} tag.
In addition to criterion-level tags, we introduce higher-level tags describing \textit{Musical Applications}, such as \textit{style transfer} or \textit{text-to-music}, to support coarse-grained filtering based on typical creative use.
The tag set is intentionally provisional and expected to evolve with future model additions and continued community discussion.

\section{Discussion}

\subsection{Insights on the Generative Music AI Landscape}

Across the evaluated systems, clear differences emerge.
Within Adaptability, only a small number of models, namely DDSP-VST \citep{ddspvst2026}, Neutone Morpho \citep{neutone2026}, and RAVE \citep{caillon2021rave}, accommodate the constraints faced by individual musicians, including modest computational requirements, support for small, personally curated datasets, and accessible adaptation pathways.
A second group of research-oriented models, including AFTER \citep{demerle2024combining}, YuE \citep{yuan2025yue}, JAM \citep{liu2025jam}, Stable Audio Open Small \citep{evans2025stableaudio}, and MusicGen \citep{copet2024simple}, provide formal mechanisms for adaptation but impose significant technical, computational, and data-related barriers, making adaptation difficult in practice.
Commercial platforms such as Suno \citep{suno2026} and Udio \citep{udio2026}, in turn, offer no practical pathways for user-driven adaptation.
By contrast, Neutone Morpho enables GPU-free model adaptation through a drag-and-drop interface, but imposes access and redistribution restrictions through platform-bound checkpoints.

Regarding Usability, a small group of systems, including DDSP-VST, Neutone Morpho, RAVE, and AFTER, show strong alignment with practical considerations such as interface availability, workflow integration, and real-time interaction, creating a clear gap with the remaining models.
By contrast, commercial platforms offer accessible interfaces and strong community support, but often restrict access, output use and ownership.
An extreme case is Udio, which prohibits downloading generated outputs.
Most of these systems also fail to support integration into music production workflows and rarely enable real-time generation for interactive or performance-oriented contexts.

A similar pattern emerges for Controllability.
The same group of systems, including DDSP-VST, Neutone Morpho, RAVE, and AFTER, provide the most extensive support for fine-grained, time-varying control over interpretable musical features, as well as multiple musically meaningful conditioning inputs.
These systems also expose a relatively rich set of control parameters, including access to internal representations that can be manipulated for creative exploration.
Other models show mixed results, typically offering coarse time-varying control or partial feature disentanglement, with text prompts as the primary conditioning mechanism.
Suno and Udio again rank lowest in this axis, providing only global forms of conditioning, largely entangled features and, in Udio's case, no explicit control parameters for shaping generative behavior.

Taken together, the evaluation reveals clear asymmetries across the three axes.
Adaptability is the most consistently challenging dimension, often constrained by technical, computational, or platform-bound factors that limit adaptation and redistribution by musicians.
Usability shows the greatest variation, with a clear separation between systems designed to integrate into existing music creation workflows and those targeting more general audiences.
This challenges democratization claims, suggesting that ease of use does not necessarily translate into meaningful support for musical practice.
Controllability occupies a more ambiguous middle ground: many systems offer partial mechanisms for steering generation, yet lack precise, time-localized control over interpretable musical features.
Overall, this points to substantial room for improvement in aligning control mechanisms with how musicians conceptualize and shape sound over time.

\subsection{A Discovery Tool for Musicians}

Beyond presenting evaluation results, the interactive MusGU+ table acts as a discovery tool that enables musicians to explore generative music systems based on concrete creative needs.
Broad \textit{Musical Application} tags support filtering for specific purposes, such as text-to-music generation or style transfer.
Filtering by evaluation axis (e.g., at least 60\% support in Adaptability, Usability, or Controllability) helps identify systems with stronger support along that particular dimension, even when musicians do not yet have clear requirements.
Reordering models by individual axes similarly supports comparison of trade-offs across dimensions.
At a finer granularity, criterion-level tags enable targeted searches, such as \textit{CPU}-only adaptation under \textit{Hardware Requirements}, \textit{VST} or \textit{Max/MSP} under \textit{Interface Availability}, or explicit \textit{timbre} control under \textit{Feature Disentanglement}.

These interaction mechanisms support exploration across different levels of abstraction, from broad use cases to more specific musical and technical requirements, without assuming a single notion of suitability.
In contrast to predominantly research-oriented model listings and benchmarks, MusGU+ foregrounds dimensions that characterize the practical conditions under which generative systems can be adopted in creative practice.

By maintaining MusGU+ as a publicly accessible and evolving resource, we aim to support continued comparison and discussion as new systems emerge.
MusGU+ is also intended to engage a broader audience across generative music research and creative practice, encouraging contributions and feedback from both perspectives.
More broadly, we hope that articulating and operationalizing these musician-centered dimensions encourages the development of generative music systems that better align with the realities of musical practice.

\subsection{Limitations and Future Directions}

A key limitation of MusGU+ is the absence of a formal user study or systematic workflow-based evaluation with musicians.
Instead, the framework is derived from prior literature, comparative system analysis, and the authors’ experience as researchers and musicians.
While appropriate for a framework-oriented contribution, this highlights the need for further experiential validation, particularly for Controllability, where future work should examine how available controls behave in practice and support fine-grained musical interaction.
By contrast, Adaptability and Usability axes rely more on observable system properties and are therefore less dependent on experiential validation at this stage.
MusGU+ also does not aim to capture experiential dimensions such as artistic or aesthetic value, creative support, or potential for collaboration, which require complementary user-centered evaluation.
Future work will incorporate workflow-based evaluations with musicians and practitioners, alongside assessment of the framework's usefulness for model discovery and selection.

A second limitation concerns the current scope of the framework.
MusGU+ focuses on audio-domain systems for general-purpose music generation and does not yet account for other paradigms, such as symbolic or MIDI-based generation, instrument-specific models, and tools that generate intermediate musical representations.
Future work will expand MusGU+ to encompass these modalities and additional models, while incorporating community contributions that help further refine the criteria and assess inter-rater agreement.

Finally, while MusGU+ does not explicitly evaluate ethical or legal dimensions, its musician-centered perspective intersects with broader concerns such as authorship, agency, cultural bias, copyright, and labor impacts, which remain unevenly addressed in the field \citep{barnett2023ethical}.
Some of these concerns relate to the dimensions considered here: limited controllability can constrain creative agency, while restricted adaptability and access can increase dependency on specific platforms.
MusGU+ is therefore not a substitute for ethical assessment, but a complementary lens that foregrounds how system design shapes musicians’ engagement with generative systems.

\section{Conclusion}

This paper introduced \textbf{MusGU+}, a musician-centered evaluation framework for generative music systems that foregrounds practical suitability for creative use.
MusGU+ provides a structured way to evaluate the practical characteristics that shape the adaptability, usability, and controllability of these systems in musical practice.

The framework is organized around three axes: Adaptability, Usability, and Controllability, and operationalized through a graded set of criteria.
We apply it to a set of representative generative music systems, revealing persistent asymmetries across these dimensions, particularly in terms of adaptability and musically meaningful control, despite improvements in accessibility.
These findings challenge claims of democratization or creative empowerment based primarily on accessibility without considering alignment with musicians' workflows and practical constraints.

We also present an interactive tool that supports exploration and comparison of systems based on musician-centered criteria and tags, enabling informed selection without reducing evaluation to a single ranking.
Thus, MusGU+ complements existing research-oriented benchmarks and frameworks by considering dimensions from a musician-centered perspective.

More broadly, MusGU+ contributes a shared vocabulary and evaluative perspective for considering generative music systems as creative tools rather than solely technical artifacts.
We hope this work encourages more reflective development practices and supports continued dialogue across technical, creative, and ethical perspectives in the design and adoption of generative music AI systems.

\section{Acknowledgments}

This work has been supported by
\textit{IA y Música: Cátedra en Inteligencia Artificial y Música} (TSI-100929-2023-1), funded by the Secretaría de Estado de Digitalización e Inteligencia Artificial, the European Union-Next Generation EU funds and BMAT Music Innovators, and
\textit{IMPA: Multimodal AI for Audio Processing} (PID2023-152250OB-I00), funded by the Ministry of Science, Innovation and Universities of the Spanish Government, the Agencia Estatal de Investigación (AEI) and co-financed by the European Union.
We thank our colleagues at the Music Technology Group, the participants of the 5th Generative Music AI Workshop, and the music producers and sound designers who contributed valuable discussion and feedback that helped refine the framework.

\bibliographystyle{apalike}
\bibliography{musgu-plus}

\appendix

\section{Detailed Evaluation Criteria}
\label{app:criteria}

\subsection{Adaptability}

Assesses how feasible it is for musicians to adapt a model to their own data, focusing on practical constraints and supported training or fine-tuning pathways.

\subsubsection{Hardware Requirements}
\begin{itemize}
    \item[\fully] The model can be trained or fine-tuned on CPU-only systems within a practical timeframe for end users.
    \item[\partially] Requires a consumer-grade GPU (e.g., T4, P100), such as those found in mid-range gaming laptops or common cloud GPU instances.
    \item[\notsupported] Requires dedicated high-end GPUs (e.g., A100s, multi-GPU rigs) or TPUs. Cannot be adapted without access to premium or institutional-level hardware.
\end{itemize}

\subsubsection{Dataset Size}
\begin{itemize}
    \item[\fully] The model can be effectively trained or fine-tuned using small, personal datasets (e.g., minutes to a few hours of audio).
    \item[\partially] The model requires a significant amount (e.g., tens of hours) of recordings or curated datasets, exceeding the size of most musicians’ own libraries.
    \item[\notsupported] The model requires large-scale datasets (e.g., hundreds of hours of diverse, high-quality material), making adaptation infeasible without access to institutional or commercial-scale data.
\end{itemize}

\subsubsection{Adaptation Pathways}
\begin{itemize}
    \item[\fully] The model provides practical pathways for training and/or fine-tuning on a musician’s own data, such as complete training code, data processing and fine-tuning scripts with required checkpoints, or a dedicated interface for this purpose.
    \item[\partially] Adaptation pathways are partially available (e.g., incomplete training code, fine-tuning code without checkpoints), requiring users to assemble or infer missing elements.
    \item[\notsupported] No practical adaptation pathways are provided. The model cannot be meaningfully trained or fine-tuned using the provided materials.
\end{itemize}

\subsubsection{Technical Barriers}
\begin{itemize}
    \item[\fully] A user-friendly graphical interface or dedicated app is provided for model adaptation, designed for musicians with no programming experience.
    \item[\partially] The model provides a streamlined setup (e.g., Colab notebook) for training or fine-tuning, with clear instructions and documentation, making it suitable for users with basic technical skills.
    \item[\notsupported] Only raw code or scripts are provided, with minimal or no guidance or documentation. Training or fine-tuning requires significant programming knowledge and low-level configuration.
\end{itemize}

\subsubsection{Model Redistribution}
\begin{itemize}
    \item[\fully] Redistribution of trained and/or fine-tuned models or checkpoints is explicitly permitted, allowing adapted models to be shared outside the original system.
    \item[\partially] Adapted models or checkpoints can be redistributed, but only under constraints (e.g., non-commercial or research-only use, platform-bound sharing).
    \item[\notsupported] Redistribution of adapted models or checkpoints is prohibited, contractually restricted, or technically impossible.
\end{itemize}

\subsection{Usability}

Examines how easily musicians can run, interact with, and integrate the model into their creative workflows, including access constraints and available support channels.

\subsubsection{Interface Availability}
\begin{itemize}
    \item[\fully] A dedicated app or user-friendly graphical interface is provided for running the model (e.g., web platform, HuggingFace Space, standalone GUI, mobile app, Max4Live device), requiring minimal or no setup.
    \item[\partially] A simplified interface (e.g., Max/MSP or Pure Data patch, Gradio UI code to be run locally) is available but requires some setup or domain-specific familiarity.
    \item[\notsupported] Only raw code or scripts are provided for inference, requiring setup from scratch and significant technical knowledge.
\end{itemize}

\subsubsection{Access Restrictions}
\begin{itemize}
    \item[\fully] The model can be used freely and repeatedly without limits, paywalls, or subscriptions. This includes open-source tools with available inference code and pretrained checkpoints, or publicly accessible platforms with no login requirements or usage limits.
    \item[\partially] Inference is possible, but access is constrained (e.g., login required, acceptance of terms of use, daily usage limits, or restricted free tiers), introducing account-based or usage limitations.
    \item[\notsupported] Direct inference is not feasible due to missing components (e.g., no pretrained checkpoint or inference code), or access is fully restricted (e.g., paywalls, subscriptions, or unstable user interfaces).
\end{itemize}

\subsubsection{Real-Time Capabilities}
\begin{itemize}
    \item[\fully] The system generates output with minimal or imperceptible delay (e.g., milliseconds to a few seconds), making it suitable for live use even on modest hardware.
    \item[\partially] Generation is possible with a moderate delay (e.g., several seconds), making it usable in interactive settings but not for live use. Real-time performance may require a consumer-grade GPU.
    \item[\notsupported] Generation is slow (e.g., minutes per sample) or impractical for real-time use, especially on typical personal computers or laptops.
\end{itemize}

\subsubsection{Workflow Integration}
\begin{itemize}
    \item[\fully] The model is directly usable in common music workflows (e.g., within DAWs, visual programming environments, live coding systems, or dedicated music hardware).
    \item[\partially] Some integration is possible (e.g., via OSC/MIDI connectivity or similar control-based interfaces).
    \item[\notsupported] The system is isolated, with no clear path to embed it in existing creative setups or musical instruments.
\end{itemize}

\subsubsection{Output Licensing}
\begin{itemize}
    \item[\fully] The output is fully usable for both personal and commercial purposes without restriction.
    \item[\partially] Some use limitations apply to the generated output (e.g., non-commercial use only, attribution required, unclear terms).
    \item[\notsupported] Output use is heavily restricted or prohibited (e.g., proprietary licensing, unclear or forbidding terms).
\end{itemize}

\subsubsection{Community Support}
\begin{itemize}
    \item[\fully] An open, user-facing community space is available (e.g., Discord server or forum), suitable for musicians to ask questions and share workflows.
    \item[\partially] Limited or developer-oriented channels are available (e.g., GitHub issues), which may provide assistance but are less accessible to most musicians.
    \item[\notsupported] No meaningful public support or community spaces are provided.
\end{itemize}

\subsection{Controllability}

Evaluates the kinds of input and internal control mechanisms a model offers to guide its behavior, including the diversity, structure, and independence of its control 
pathways.

\subsubsection{Conditioning Inputs}
\begin{itemize}
    \item[\fully] The model accepts multiple and diverse conditioning inputs, including at least two musically meaningful modalities (e.g., audio, MIDI, symbolic), enabling rich and varied guidance during generation.
    \item[\partially] Conditioning is limited to one musically meaningful modality, possibly combined with descriptive inputs (e.g., melody and text).
    \item[\notsupported] Offers little to no input conditioning. Generation is mostly uncontrolled or limited to coarse or global labels (e.g., vibe, energy).
\end{itemize}

\subsubsection{Time-Varying Control}
\begin{itemize}
    \item[\fully] Supports precise time-varying control (e.g., per-frame pitch, dynamics, or symbolic structure), enabling fine-grained and expressive manipulation.
    \item[\partially] Some time-localized control is available (e.g., melody following or segment-based prompts), but not fine-grained.
    \item[\notsupported] Only global control is possible, with no ability to influence generation over time.
\end{itemize}

\subsubsection{Feature Disentanglement}
\begin{itemize}
    \item[\fully] Provided controls are designed to influence distinct and isolated musical attributes (e.g., timbre, pitch, rhythm, structure), enabling predictable and interpretable guidance.
    \item[\partially] Some effort is made to separate control pathways, but conditioning inputs are not explicitly associated with interpretable musical attributes, and interactions are expected.
    \item[\notsupported] Control signals are entangled or loosely defined, making it unclear how specific inputs relate to individual musical attributes.
\end{itemize}

\subsubsection{Control Parameters}
\begin{itemize}
    \item[\fully] Provides multiple configurable model parameters (e.g., duration, randomness, conditioning strength) and enables direct manipulation of internal representations.
    \item[\partially] Provides a limited set of configurable parameters and/or restricted access to latent representations.
    \item[\notsupported] No additional meaningful parameters or internal controls are exposed beyond the primary conditioning mechanisms, if any.
\end{itemize}

\end{document}